\documentclass[twocolumn]{aastex63}

\usepackage{amsmath}

\begin{document}

\title{The time delay distribution and formation metallicity of LIGO-Virgo's binary black holes}

\author{Maya Fishbach}
\altaffiliation{NASA Hubble Fellowship Program Einstein Postdoctoral Fellow}
\affiliation{Center for Interdisciplinary Exploration and Research in Astrophysics (CIERA) and Department of Physics and Astronomy,
Northwestern University, 1800 Sherman Ave, Evanston, IL 60201, USA}

\author{Vicky Kalogera}
\affiliation{Center for Interdisciplinary Exploration and Research in Astrophysics (CIERA) and Department of Physics and Astronomy,
Northwestern University, 1800 Sherman Ave, Evanston, IL 60201, USA}

\begin{abstract}
We derive the first constraints on the time delay distribution of binary black hole (BBH) mergers using the LIGO-Virgo Gravitational-Wave Transient Catalog GWTC-2. Assuming that the progenitor formation rate follows the star formation rate (SFR), the data favor that $43$--$100\%$ of mergers have delay times $<4.5$ Gyr (90\% credibility). Adopting a model for the metallicity evolution, we derive joint constraints for the metallicity-dependence of the BBH formation efficiency and the distribution of time delays between formation and merger. Short time delays are favored regardless of the assumed metallicity dependence, although the preference for short delays weakens as we consider stricter low-metallicity thresholds for BBH formation. For a $p(\tau) \propto \tau^{-1}$ time delay distribution and a progenitor formation rate that follows the SFR without metallicity dependence, we find that $\tau_\mathrm{min}<2.2$ Gyr, whereas considering only the low-metallicity $Z < 0.3\,Z_\odot$ SFR, $\tau_\mathrm{min} < 3.0$ Gyr (90\% credibility). Alternatively, if we assume long time delays, the progenitor formation rate must peak at higher redshifts than the SFR. For example, for a $p(\tau) \propto \tau^{-1}$ time delay distribution with $\tau_\mathrm{min} = 4$ Gyr, the inferred progenitor rate peaks at \replaced{$z = 5.4^{+3.0}_{-3.2}$ (90\% credible interval)}{$z > 3.9$ (90\% credibility)}. Finally, we explore whether the inferred formation rate and time delay distribution vary with BBH mass.
\end{abstract}

\section{Introduction}
\label{sec:intro}
The latest catalog of compact binary coalescences observed by Advanced LIGO~\citep{2015CQGra..32g4001L} and Virgo~\citep{2015CQGra..32b4001A}, GWTC-2, includes BBH mergers out to $z \sim 1$~\citep{2020arXiv201014527A,2020arXiv201014533T}. These observations probe the evolution of the BBH population over the last $\sim8$ billion years, providing updated constraints on the merger rate~\citep{2020arXiv201014533T,2021arXiv210112130T} and the mass distribution~\citep{2021arXiv210107699F} as a function of redshift.

Measuring the rate of BBH mergers as a function of redshift yields valuable clues to their evolutionary histories. The BBH merger rate depends on a combination of the progenitor formation rate and the distribution of delay times between formation and merger~\citep{2010ApJ...716..615O,2010MNRAS.402..371B,2013ApJ...779...72D,2016MNRAS.458.2634M,2016Natur.534..512B,2016MNRAS.463L..31L,2018PhRvL.121p1103F,2018MNRAS.481.1908K,2018MNRAS.473.1186E,2018ApJ...866L...5R,2019MNRAS.490.3740N,2019MNRAS.482.5012C, 2020ApJ...898..152S,2020MNRAS.499.5941D,2020MNRAS.493L...6T,2021MNRAS.502.4877S}. 

In most formation channels, black holes have a stellar origin, and the progenitor formation rate is closely related to the SFR~\citep{2014ARA&A..52..415M,2015MNRAS.447.2575V,2017ApJ...840...39M,2019MNRAS.482.4528E}. Because the BBH formation efficiency is expected to be a strong function of the stellar metallicity~\citep{2000ARA&A..38..613K,2010ApJ...715L.138B, 2011A&A...530A.115B,2012ApJ...749...91F}, the progenitor formation rate also depends on the metallicity evolution of the universe~\citep{2006ApJ...638L..63L,2015MNRAS.453..960M,2019MNRAS.488.5300C,2020A&A...636A..10C}.

Meanwhile, the time delay between formation and merger is a unique property of the formation channel, and different proposed channels predict different distributions of time delays. Classical isolated binary evolution predicts a power-law time delay distribution $p(\tau) \propto \tau^\alpha$, dominated by the GW merger timescale $t \propto a^4$ for an initial orbital separation $a$~\citep{1964PhRv..136.1224P}. A typical prediction for the power-law slope is $\alpha = -1$~\citep{2010ApJ...716..615O,2012ApJ...759...52D}, assuming a flat-in-log distribution for the initial binary separations $p(a) \propto a^{-1}$~\citep{1983ARA&A..21..343A,2013A&A...550A.107S}, but the exact form of the distribution depends on uncertain physics including stellar winds, mass transfer, and kicks, in addition to the uncertain distribution of orbital separations~\citep{2008ApJ...675..566O,2010ApJ...716..615O,2017MNRAS.472.2422M}.
A subclass of isolated binary evolution, chemically homogeneous evolution predicts longer delay times~\citep{2016MNRAS.458.2634M,2016A&A...588A..50M}, with a strong correlation between the formation metallicity and the delay time to merger. At the highest metallicities possible for this channel, $Z \sim 0.2\,Z_\odot$, the predicted delay times are long, $\tau > 3.5\,\mathrm{Gyr}$~\citep{2016MNRAS.458.2634M}, whereas at much lower metallicities ($Z \lesssim 0.01\,Z_\odot$), shorter delay times $\tau < 1\,\mathrm{Gyr}$ are possible~\citep{2016A&A...588A..50M,2020MNRAS.499.5941D}. 
Stellar evolution in triple stellar systems, rather than binaries, may also produce BBH mergers, with a time delay distribution skewed toward longer time delays than the isolated binary case, with most delays larger than 1 Gyr~\citep{2017ApJ...841...77A,2018ApJ...863....7R,2018ApJ...856..140H}.
For BBH formation in young star clusters, most systems experience short delay times, with the predicted delay time distribution peaking at $\sim 100$ Myr and following a $\tau^{-1}$ distribution above 400 Myr~\citep{2020MNRAS.498..495D}.
For dynamically assembled BBHs in globular clusters, the delay time distribution depends on the cluster's virial radius, with larger radii leading to larger delay times~\citep{2018PhRvL.120o1101R}. BBH systems ejected from the cluster prior to merger ($\sim 50\%$ of BBH mergers) are expected to experience very long delay times (with a median of $\sim10$ Gyr), while in-cluster mergers occur extremely promptly, which may lead to trends between eccentricity, mass, and spin with merger redshift~\citep{2013LRR....16....4B,2016PhRvD..93h4029R,2017MNRAS.467..524B,2018PhRvL.120o1101R,2018PhRvD..97j3014S,2020ApJS..247...48K,2021MNRAS.503.3371B}. Another proposed site for the dynamical assembly of BBHs is the disks of active galactic nuclei (AGN), which are expected to merge BBHs with short delay times $\tau < 100$ Myr~\citep{2020ApJ...896..138Y}. However, the BBH merger rate in AGN is expected to peak at relatively low redshifts $z \lesssim 1$ because their formation traces the evolution of the AGN luminosity function~\citep{2020ApJ...896..138Y}.

Previous work has demonstrated that GW observations can meaningfully constrain the evolution of the merger rate by using a catalog of LIGO-Virgo BBH events at $z \lesssim 1.5$ ~\citep{2018ApJ...863L..41F,2019ApJ...882L..24A,2020arXiv201014533T,2020PhRvD.102l3022R,2020arXiv201208839T}, combining a BBH catalog with a (non)detection of the astrophysical stochastic background~\citep{2020ApJ...896L..32C,2020ApJ...901..137S,2021arXiv210112130T}, or anticipating the next generation of ground-based GW detectors, which would trace the evolution of the merger rate across the entire observable universe~\citep{2019ApJ...886L...1V,2019ApJ...878L..13S,2019BAAS...51c.242K,2020arXiv201209876N,2020arXiv201114541R}.

In this work, we derive the first observational constraints on the BBH time delay distribution, the progenitor formation rate, and its metallicity dependence. We describe a phenomenological fit to the redshift evolution of the BBH merger rate in Section~\ref{sec:z-evol}. Fixing the progenitor formation rate to the (low-metallicity) SFR, we derive constraints on the time delay distribution in Section~\ref{sec:td}. In Section~\ref{sec:formation}, we measure the threshold metallicity for BBH formation and the corresponding progenitor formation rate for fixed time delay distributions. We also explore how the formation rate and time delay may depend on the mass of the BBH system in Section~\ref{sec:mass}, and discuss the implications of these findings for BBH formation scenarios in Section~\ref{sec:conclusion}. Although the GWTC-2 events only probe redshifts $z \lesssim 1$ far below the peak of the SFR at $z \sim 2$, we find that we can already derive useful astrophysical constraints from the observed redshift evolution.

\section{BBH merger rate as a function of redshift}
\label{sec:z-evol}
We begin by reviewing the inferred BBH merger rate from GWTC-2. As in~\citet{2020arXiv201014533T,2021arXiv210112130T}, we jointly fit the mass, spin, and redshift distribution of the BBH population with simple phenomenological models described in Appendix~\ref{sec:methods}. The inferred redshift evolution has significant correlations with the mass distribution~\citep{2018ApJ...863L..41F}, and we discuss possible systematic uncertainties associated with the choice of mass model in Section~\ref{sec:mass}. In our calculation, we do not take into account the stochastic GW background upper limit reported in~\citet{2021arXiv210112130T} because at this stage, it does not provide much additional information compared to the resolved BBH events. Thus, we only consider the merger rate up to $z = 1$. However, like~\citet{2021arXiv210112130T}, our redshift model allows the merger rate to peak at some redshift $z_p$, finding that the data disfavor $z_p < 1$.

\begin{figure}
    \centering
    \includegraphics[width = 0.5\textwidth]{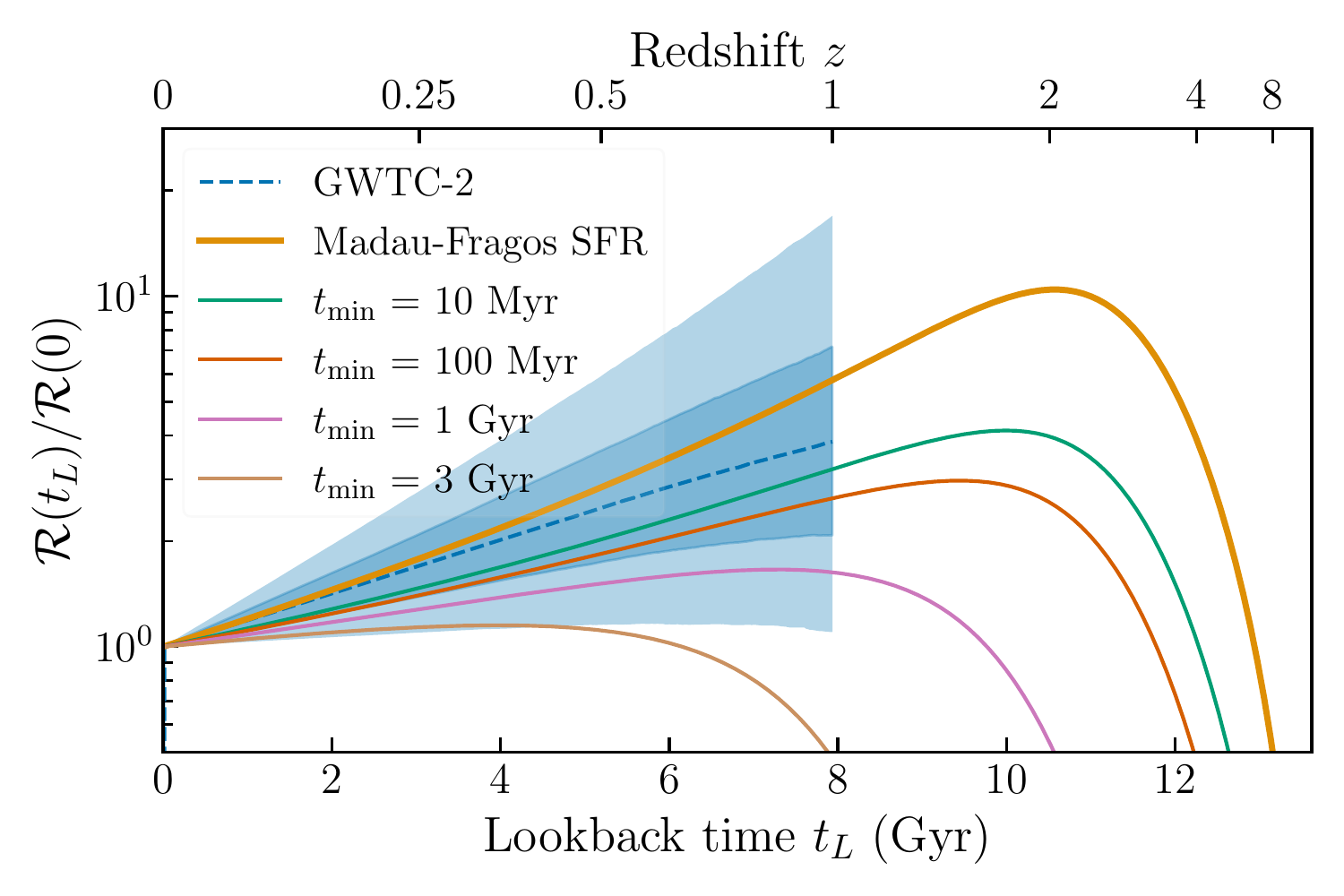}
    \caption{Merger rate as a function of lookback time (or equivalently, redshift) normalized to the merger rate today. In blue, evolution of the merger rate inferred from fitting a population model to the GWTC-2 events. The dashed blue line shows the median, while the shaded bands show 50\% and 90\% credible intervals. For comparison, the thick gold line shows the Madau-Fragos SFR. The green, orange, pink, and tan lines illustrate predicted merger rate shapes, assuming a formation rate that follows the Madau-Fragos SFR and a $\tau^{-1}$ time delay distribution with a minimum time delay of 50 Myr, 500 Myr, 1 Gyr and 3 Gyr respectively. If the progenitor formation rate follows the SFR, a distribution of time delays that peaks at long times $\tau \gtrsim 3$ Gyr is in tension with the inferred distribution from GWTC-2.}
    \label{fig:RofTime}
\end{figure}

Figure~\ref{fig:RofTime} shows the merger rate evolution $\mathcal{R}(z)$ inferred by fitting the phenomenological $p(m_1, m_2, \chi_\mathrm{eff}, z)$ model described in Appendix~\ref{sec:methods} to the 44 confident GWTC-2 BBH events analyzed in~\citet{2020arXiv201014533T}. The blue bands show 50\% and 90\% symmetric credible regions, while the dashed blue line shows the median.
For comparison, we show example merger rate curves for different time delay models that follow $p(\tau) \propto \tau^{-1}$ with different minimum time delays $\tau_\mathrm{min}$. We assume that the progenitor formation rate follows the \citet{2017ApJ...840...39M} SFR. In the following sections, we also consider progenitor formation rates that follow the \emph{low-metallicity} SFR, for some $Z < Z_\mathrm{thresh}$, rather than the \emph{total} SFR, adopting the mean metallicity-redshift relation from~\citet{2017ApJ...840...39M}. The calculation of the merger rate $\mathcal{R}(z)$ given the progenitor formation rate and the time delay distribution is detailed in Appendix~\ref{sec:methods}. From Figure~\ref{fig:RofTime}, we see that the GWTC-2 measurement of the rate evolution, in reference to the SFR, is informative about the time delay distribution, disfavoring distributions with large $\tau_\mathrm{min} \gtrsim 3$ Gyr.

\section{Time delay inference}
\label{sec:td}
The previous section showed a phenomenological fit to the merger rate evolution $\mathcal{R}(z)$. In this section, we adopt a physical parameterization for the redshift evolution by modeling the progenitor formation rate $R_f(z)$ and the time delay distribution $p(\tau)$ (see Eq.~\ref{eq:rate-integral}). Throughout, we consider maximum time delays of $\tau = 13.5$ Gyr, corresponding to maximum formation redshifts of $z = 14$. Fixing the progenitor formation rate to the SFR, we fit for the time delay distribution under a binned histogram model (Section~\ref{sec:binned}) and a power-law model (Section~\ref{sec:pl-timedelay}). We then consider progenitor rates that follow the low-metallicity SFR, and explore how the metallicity-dependence affects the time delay inference (Section~\ref{sec:metallicity}). In Section~\ref{sec:mass} we investigate possible correlations between BBH mass and delay times.

\subsection{Binned time delay model}
\label{sec:binned}
We begin by modeling the time delay distribution as piecewise constant in $n$ bins with bin edges given by $\{b_i\}_{i = 1}^{n+1}$, 
\begin{equation}
\label{eq:binned}
    p(\tau) = \sum_{i = 1}^{n} p_i \Theta(b_i \leq \tau < b_{i+1}),
\end{equation}
where $\Theta$ is an indicator function.

\begin{figure}
    \centering
    \includegraphics[width=0.5\textwidth]{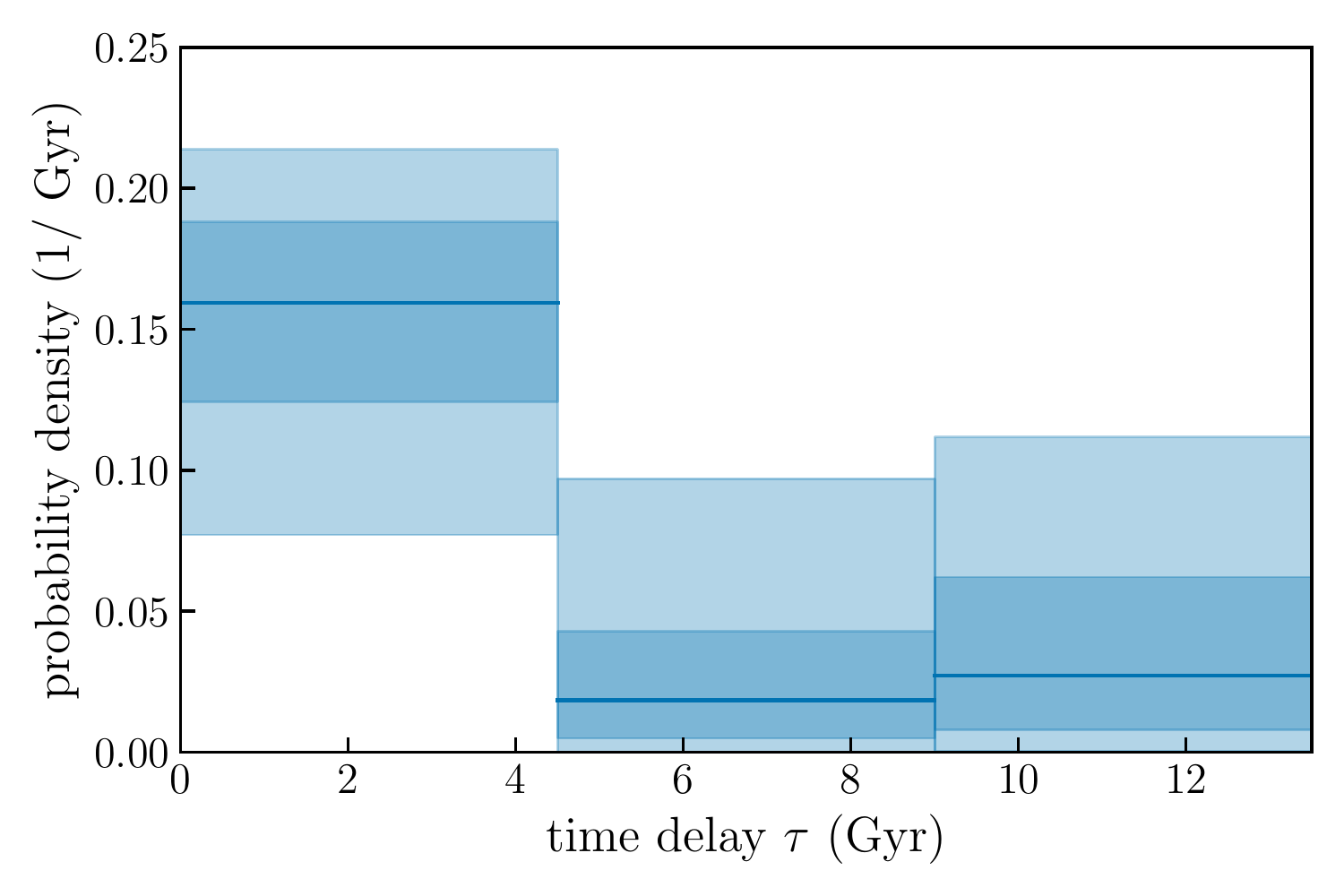}
    \caption{Inferred time delay distribution assuming a binned time delay model for a progenitor rate that follows the SFR. The data are consistent with all BBH systems having time delays smaller than 4.5 Gyr (belonging to the first bin), and requires that at least 43\% of mergers experience time delays smaller than 4.5 Gyr.}
    \label{fig:binnedTD}
\end{figure}

We consider three equally spaced time bins up to $\tau_\mathrm{max} = 13.5$ Gyr, fixing the bin edges $b_i = \frac{13.5(i - 1)}{3}$ Gyr. We obtain qualitatively similar results when we consider a five-binned model, with larger uncertainties on the bin heights as expected for a model with more free parameters.

As in Section~\ref{sec:z-evol}, we fit the joint mass-redshift-spin distribution of the BBH population, but we replace the rate evolution model with this physical parameterization.
We use priors on the mass and spin parameters as in Appendix~\ref{sec:methods}. For the redshift distribution, we choose the Jeffreys prior on the fraction $p_i \times (b_{i+1} - b_i)$ of systems within each time delay bin. The Jeffreys prior for a trinomial distribution is a Dirichlet distribution with concentration parameters $\alpha = 0.5$~\citep{categoricalstats}.
We fix the shape of the formation rate $R_f(z)$ to follow the Madau-Fragos SFR.

 The inferred time delay distribution $p(\tau)$ is shown in Figure~\ref{fig:binnedTD}. The data are consistent with all mergers belonging to the smallest time delay bin. This is true regardless of the bin boundaries; because the shape of the merger rate evolution is consistent with the SFR, the data are consistent with all BBH mergers having arbitrarily small time delays. The data requires that 43-100\% of delay times are smaller than 4.5 Gyr (90\% credibility).
%We recover similar results when we consider a formation rate that follows the low-metallicity $Z < 0.3\,Z_\odot$ SFR (assuming a 0.4 dex spread in the metallicity-redshift relation); in this case the data requires that 37-100\% of time delays fall in the first bin with $\tau < 4.5$ Gyr. 
While there is a preference for small time delays, this flexible model permits a broad range of time delay distributions. In the following section we explore stricter parameterizations for the time delay distribution.

\subsection{Power-law time delay model}
\label{sec:pl-timedelay}
\begin{figure*}
    \centering
    \includegraphics[width = \textwidth]{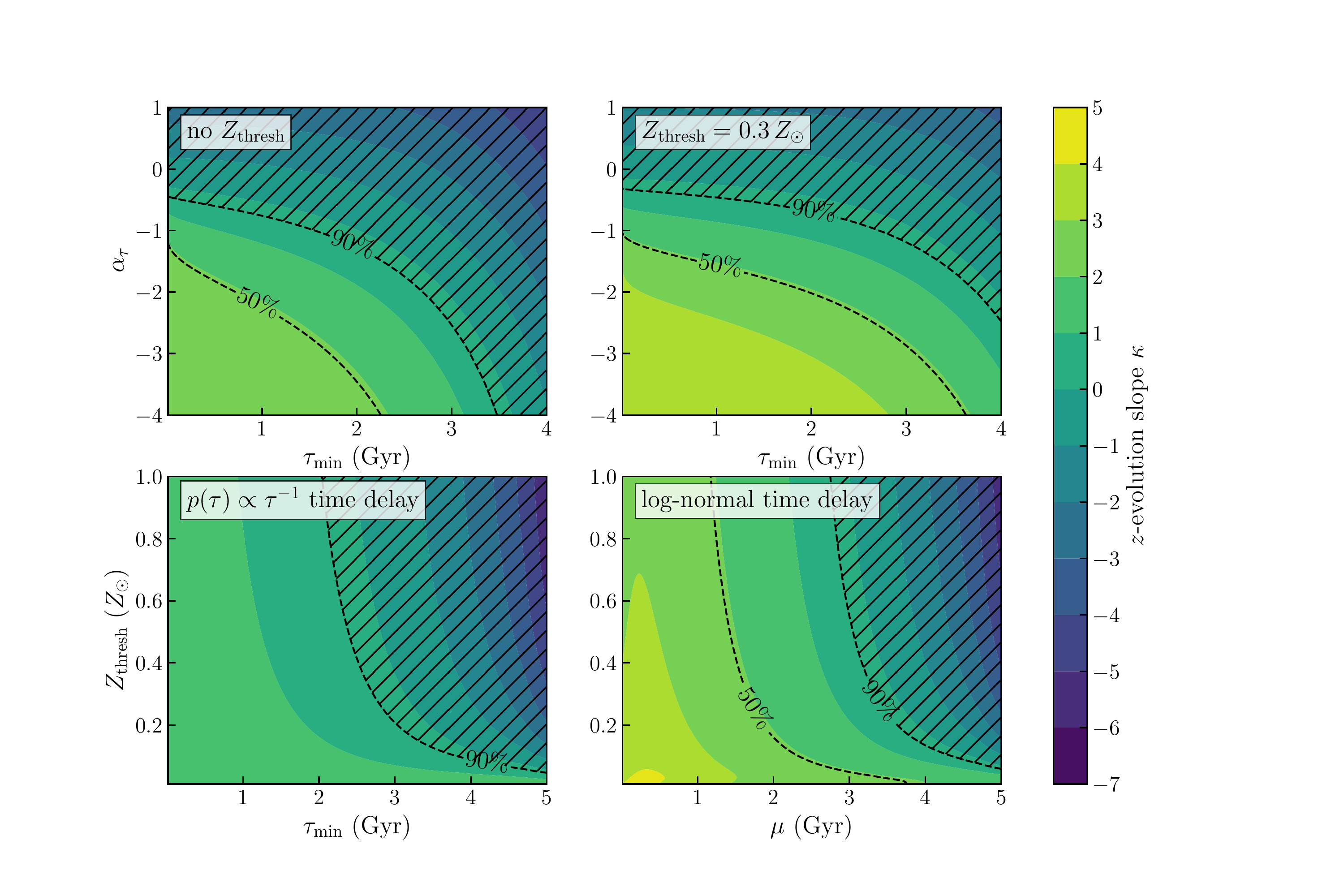}
    \caption{Approximate redshift evolution parameter $\kappa$ where $\mathcal{R}(z) = R_0(1+z)^\kappa$ resulting from different time delay and progenitor formation models (filled contours). The dashed black contours denote the 50\% and 90\% credible bounds on the rate evolution from the GWTC-2 inference, so that the black hatched region of parameter space is excluded at 90\% credibility. \emph{Top left:} The formation rate is assumed to follow the Madau-Fragos SFR, and the time delay distribution follows a power law with variable $\alpha_\tau$ and $\tau_\mathrm{min}$. \emph{Top right:} Same as the top left panel, but the formation rate is assumed to follow the low-metallicity SFR with $Z_\mathrm{thresh} = 0.3\,Z_\odot$. We assume a 0.4 dex spread in the mean metallicity-redshift relation of~\citet{2017ApJ...840...39M} (Eq.~\ref{eq:Z-z}). \emph{Bottom left:} We vary the threshold metallicity and the minimum time delay in a $\tau^{-1}$ time delay distribution. The inferred $\tau_\mathrm{min}$ is relatively independent of $Z_\mathrm{thresh}$ as long as $Z_\mathrm{thresh} \gtrsim 0.2\,Z_\odot$. Repeating the calculation with the mean metallicity-redshift relation of~\citet{2006ApJ...638L..63L} yields similar results, with the resulting values for $\kappa$ increasing by a nearly constant amount of $\sim 0.3$ across the parameter space. \emph{Bottom right:} Same as bottom left, but rather than a $\tau^{-1}$ time delay distribution, we assume a log-normal time delay distribution (Eq.~\ref{eq:lognorm}) centered at $\mu$ with a width of 0.5 dex.}
    \label{fig:kappa_timedelay}
\end{figure*}

%\begin{figure}
%    \centering
%    \includegraphics[width = 0.5\textwidth]{Rz1z0_ratio_timedelay_MFSFR.pdf}
%    \caption{Expected ratio of the merger rate at $z = 1$ to $z = 0$, $\mathcal{R}(z = 1)/ \mathcal{R}(z=0)$ as a function of the time delay distribution in a power-law time delay model (filled contours). The formation rate is assumed to follow the Madau-Fragos SFR. The dashed white contours denote the 50\% and 90\% credible bounds on the rate evolution from the GWTC-2 inference.}
%    \label{fig:rateratio_timedelay}
%\end{figure}

\begin{figure}
    \centering
    \includegraphics[width = 0.5\textwidth]{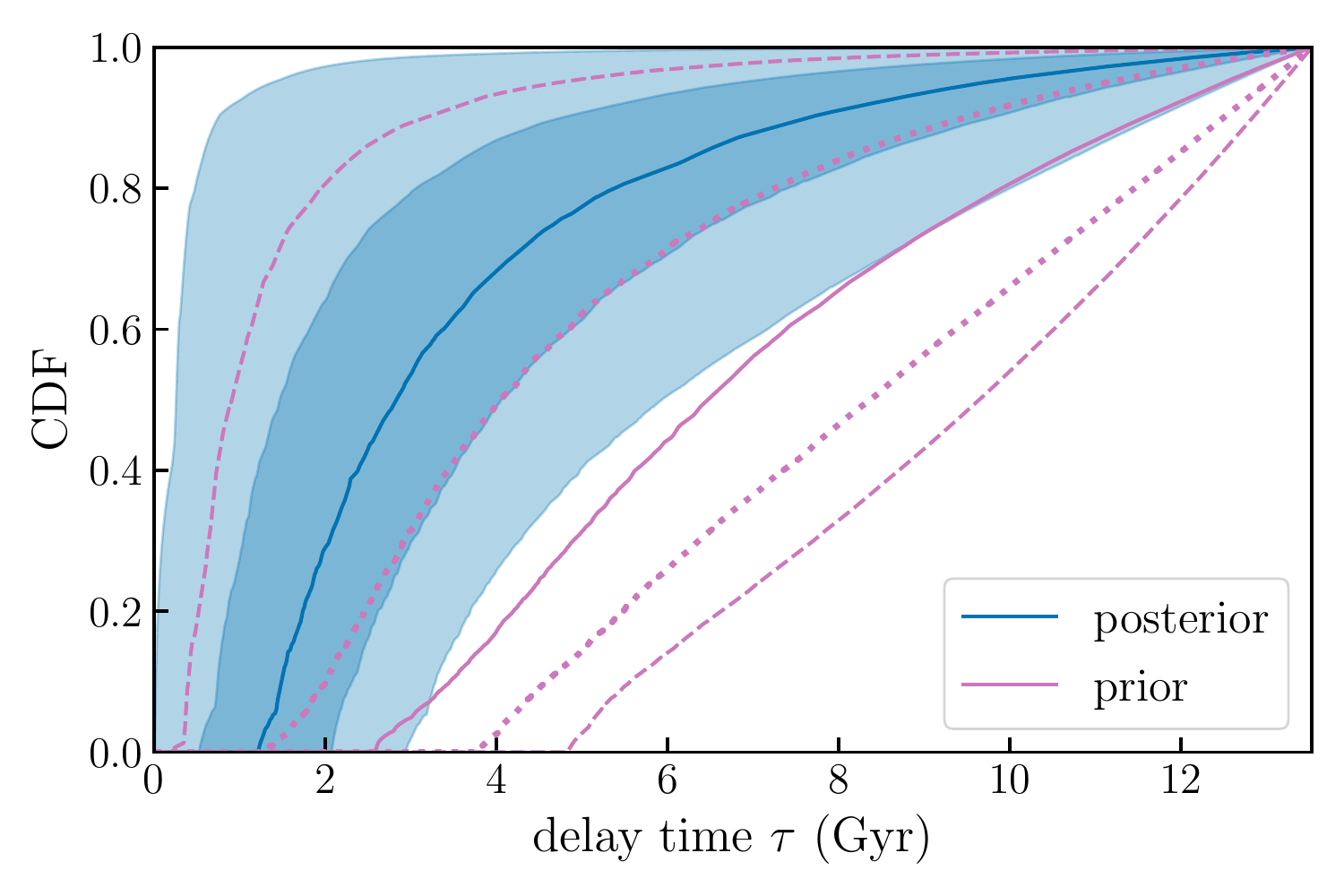}
    \caption{Inferred time delay distribution, shown as the cumulative distribution function (CDF) for a power-law model, assuming that the formation rate follows the SFR. Solid lines show the median CDF \added{for the posterior (blue) and the prior (pink)}. Shaded blue bands enclose the central 50\% and 90\% of posterior CDFs at each delay time. \added{The pink dotted (dashed) lines enclose 50\% (90\%) of the prior CDFs.} Compared to the prior, the posterior favors short time delays.}
    \label{fig:CDF_PLtimedelay}
\end{figure}

In this subsection, we consider a power-law model for the time delay distribution, parameterized by a slope $\alpha_\tau$ and a minimum delay time $\tau_\mathrm{min}$:
\begin{equation}
\label{eq:ptau-pl}
    p(\tau) \propto \tau^{\alpha_\tau} \Theta(\tau_\mathrm{min} \leq \tau < 13.5\, \mathrm{Gyr}).
\end{equation}
Because the current GW catalog extends only to $z = 1$, we find that the information about the time delay distribution can be summarized by the inferred merger rate at $z = 0$ compared to $z = 1$, or $\mathcal{R}(z = 1) \equiv \mathcal{R}_1$ compared to $\mathcal{R}(z = 0) \equiv \mathcal{R}_0$. Any combination of progenitor formation rate $R_f$ and time delay distribution predicts a value for the ratio $\mathcal{R}_1/\mathcal{R}_0$. We map this quantity onto an effective $(1 + z)$ power-law slope parameter $\kappa$, taking $\kappa = \ln \left(\mathcal{R}_1/\mathcal{R}_0 \right) / \ln 2$, and summarize the time delay inference in terms of $\kappa$. 

Figure~\ref{fig:kappa_timedelay} shows the effective $\kappa$ parameter for a family of time delay distributions and progenitor formation rates. In the top left panel (``no $Z_\mathrm{thesh}$''), we fix the formation rate to the SFR and consider power-law time delay distributions characterized by a power-law slope $\alpha_\tau$ and a minimum time delay $\tau_\mathrm{min}$. Steeper (more negative) power-law slopes imply larger values of $\kappa$, as do smaller values of the minimum time delay.
We overlay dashed black contours corresponding to the 50\% and 90\% credible intervals on the merger rate evolution $\mathcal{R}(z)$ inferred from GWTC-2 in Section~\ref{sec:z-evol}. The model parameter space outside the 90\% contour (the hatched region) is ruled out by GWTC-2 at 90\% credibility. We can see from the top left panel of Figure~\ref{fig:kappa_timedelay} that in order to match the inferred redshift evolution between $z = 0$ and $z = 1$, the time delay distribution must be relatively steep ($\alpha_\tau \lesssim -0.5$) and/or peak at a small minimum time delay ($\tau_\mathrm{min} \lesssim 3.5$ Gyr).

Fixing the shape of the progenitor formation rate to the SFR (equivalently, fixing $Z_\mathrm{thresh}$ to be large) and simultaneously fitting for the SFR normalization $R_{f,0}$ and $(\alpha_\tau, \tau_\mathrm{min})$ in the power-law time delay model of Eq.~\ref{eq:ptau-pl}, we obtain the posterior on the distribution of time delays shown in Figure~\ref{fig:CDF_PLtimedelay}. This figure shows the inference on the cumulative distribution function (CDF) of time delays for the posterior (blue) compared to the prior (pink). Under this model, we obtain median delay times $\tau_\mathrm{50\%} = 2.8^{+3.3}_{-2.6}$ Gyr, compared to the prior $\tau_\mathrm{50\%} = 6.5^{+3.1}_{-5.6}$ Gyr (90\% symmetric credible interval). In the following subsection we will explore how restricting to the low-metallicity SFR affects these results. 

\subsection{Effect of metallicity}
\label{sec:metallicity}

%\begin{figure}
%    \centering
%    \includegraphics[width = 0.5\textwidth]{Rz1z0_ratio_timedelay_metallicity_sigmaz04.pdf}
%    \caption{Approximate redshift evolution parameter $\kappa$, as in Figure~\ref{fig:kappa_timedelay}, as a function of the minimum time delay for a flat-in-log time delay distribution and the threshold metallicity for progenitor formation. The dashed black contour corresponds to the 90\% posterior bound on the merger rate evolution. We assume a 0.4 dex spread in the mean metallicity-redshift relation of~\citet{2017ApJ...840...39M} (Eq.~\ref{eq:Z-z}). Repeating the calculation with the mean metallicity-redshift relation of~\citet{2006ApJ...638L..63L} yields similar results, with the resulting values for $\kappa$ increasing by a nearly constant amount of $\sim 0.3$ across the parameter space.}
%    \label{fig:kappa_metallicity}
%\end{figure}

%\begin{figure}
%    \centering
%    \includegraphics[width = 0.5\textwidth]{CDF_PLtimedelay_varyZthresh_04dex.pdf}
%    \caption{Inferred time delay distribution for a power-law model, assuming that the formation rate follows the SFR (blue) or the low-metallicity SFR with $Z < 0.3\,Z_\odot$ (orange) and $Z < 0.1\,Z_\odot$ (green). The solid lines denote the median and shaded bands denote 90\% credible intervals. We assume a scatter around the mean metallicity-redshift relation of 0.4 dex.}
%    \label{fig:CDF_PLtimedelay_varyZ}
%\end{figure}

\begin{figure}
    \centering
    \includegraphics[width = 0.5\textwidth]{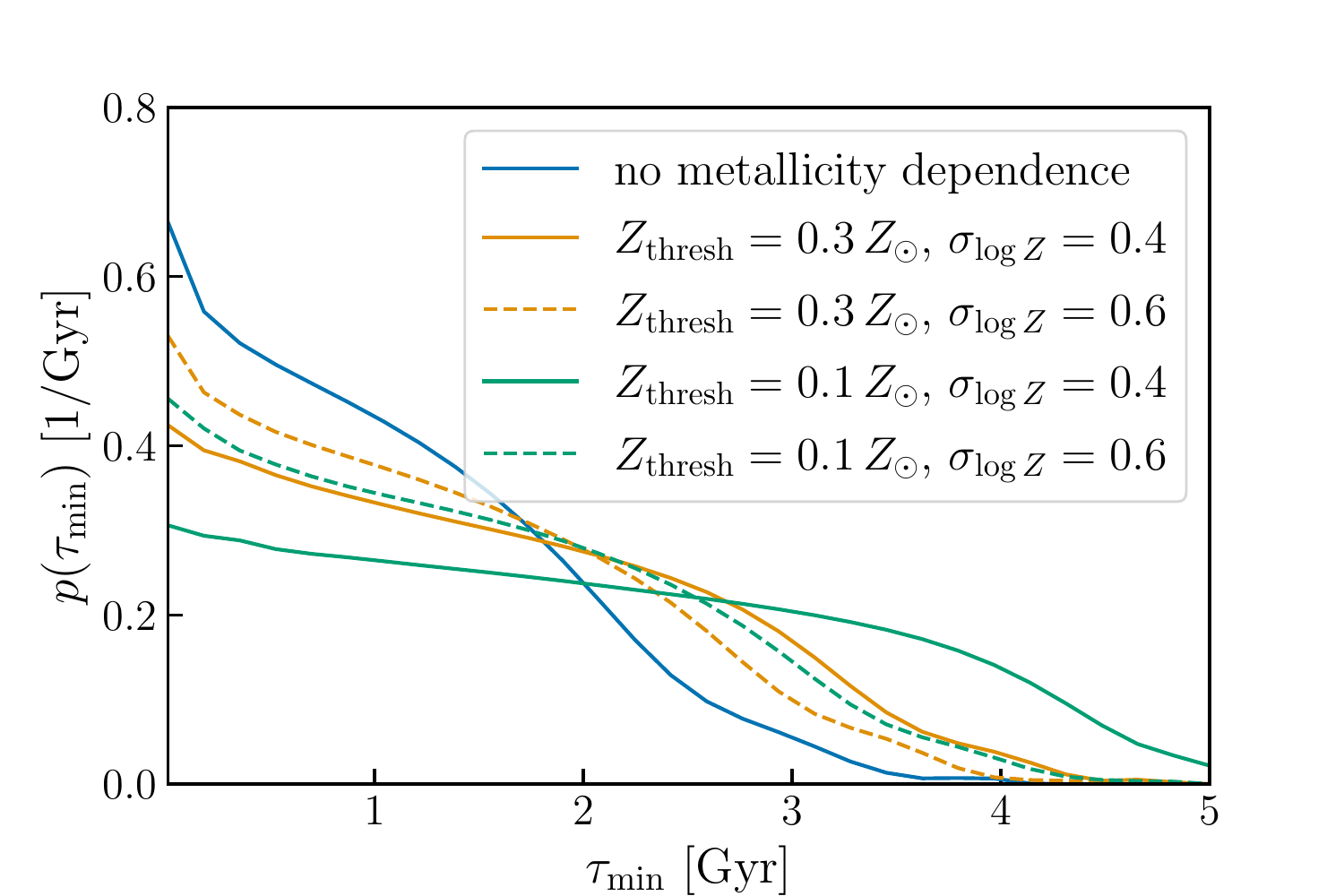}
    \caption{Posterior distribution of the minimum time delay $\tau_\mathrm{min}$ for a $p(\tau) \propto \tau^{-1}$ time delay distribution. The different lines correspond to different assumptions about the metallicity-dependence of the progenitor formation rate and the distribution of metallicities at a given redshift.}
    \label{fig:ptau-min_metallicity}
\end{figure}

In this subsection we infer time delay distributions corresponding to progenitor formation rates that follow the low-metallicity SFR, parameterized by a threshold metallicity $Z_\mathrm{thresh}$. We assume a default value for the scatter about the mean metallicity-redshift relation of $\sigma_{\log Z}$ = 0.4 dex.

%Comparing the top right panel of Figure~\ref{fig:kappa_timedelay} (``$Z_\mathrm{thresh} = 0.3\,Z_\odot$'') to the top left panel (``no $Z_\mathrm{thresh}$'') discussed in the previous subsection, we see how the metallicity threshold affects the merger rate evolution for the same time delay distributions. 
We simultaneously vary $Z_\mathrm{thresh}$ and the time delay distribution in the bottom two panels of Figure~\ref{fig:kappa_timedelay}. The bottom left panel (``$p(\tau) \propto \tau^{-1}$ time delay") shows the joint effect of varying $Z_\mathrm{thresh}$ and $\tau_\mathrm{min}$ on the merger rate evolution for a $\tau^{-1}$ time delay model. The bottom right panel (``log-normal time delay"), instead of a $\tau^{-1}$ time delay model, uses a truncated log-normal:
\begin{align}
\label{eq:lognorm}
    p(\tau) \propto &\tau^{-1}\exp\left[{-\frac{1}{2}\left(\frac{\log_{10} \tau - \log_{10} \mu}{s}\right)^2}\right] \times \nonumber \\ &\Theta(0 \leq \tau < 13.5),
\end{align}
with $\mu$ shown on the x-axis and $s$ fixed to $s = 0.5$ dex. 

As we consider a stricter metallicity cut (smaller $Z_\mathrm{thresh}$ values) for BBH formation, we move the peak of the progenitor formation rate to higher redshifts, and the data correspondingly allow for larger $\tau_\mathrm{min}$ or $\mu$ to match the observed redshift evolution. 
Nevertheless, we can see from the bottom two panels of Figure~\ref{fig:kappa_timedelay} that the preference for small delay times $\tau \lesssim 3$ Gyr persists across all $Z_\mathrm{thresh}$.

%For example, by comparing the top two panels of Figure~\ref{fig:kappa_timedelay}, we can see that for ``no $Z_\mathrm{thresh}$" as well as for ``$Z_\mathrm{thresh} = 0.3\,Z_\odot$", the joint posterior on the time delay power law parameters $(\alpha_\tau, \tau_\mathrm{min})$ favor the smallest $\alpha_\tau$ and $\tau_\mathrm{min}$ considered. 
We can also repeat the analyses of the previous two subsections fixing the progenitor formation rate to the low-metallicity SFR with $Z < 0.3\,Z_\odot$, rather than the total SFR; see the top right panel (``$Z_\mathrm{thresh} = 0.3\,Z_\odot$") of Figure~\ref{fig:kappa_timedelay}.
When we assumed that the progenitor formation followed the SFR with no metallicity dependence, we found that 43--100\% of systems experience delay times under 4.5 Gyr in the binned model and a median delay time of $\tau_{50\%} = 2.8^{+3.3}_{-2.6}$ Gyr in the power-law model. If we instead assume that the progenitor formation rate follows the low-metallicity SFR with $Z_\mathrm{thresh} = 0.3\, Z_\odot$ and repeat these analyses, we find that 37--100\% of systems experience delay times under 4.5 Gyr in the binned model, and that in the power-law model, the inferred median delay time is $\tau_\mathrm{50\%} = 3.9^{+3.2}_{-3.5}$ Gyr. With an even stricter metallicity threshold of  $Z_\mathrm{thresh} = 0.1\, Z_\odot$, the inferred median delay time is $\tau_\mathrm{50\%} = 4.9^{+3.0}_{-4.3}$ Gyr. Within statistical uncertainties, the inferred time delay distributions are consistent across different values of $Z_\mathrm{thresh}$, suggesting that the assumed metallicity threshold does not strongly impact the conclusions about the time delay distribution.

%We vary the threshold metallicity as well as the scatter around the mean metalllicity-redshift relation of Eq.~\ref{eq:Z-z}. The resulting inference for the time delay distribution is summarized in Figures~\ref{fig:kappa_metallicity} and~\ref{fig:ptau-min_metallicity}.

%Figure~\ref{fig:CDF_PLtimedelay_varyZ} shows the cumulative time delay distribution for different values of the threshold metallicity for BBH formation, $0.3\,Z_\odot$ and $0.1\,Z_\odot$. For this plot, we assume a scatter around the mean metallicity-redshift relation of 0.4 dex, which is the default scatter assumed in~\citet{2019MNRAS.490.3740N,2021arXiv210212495B}. As we consider a stricter metallicity cut for BBH formation, the peak of the low-metallicity star formation moves to higher redshifts and the data correspondingly allow for longer time delays to match the observed redshift evolution. \maya{Include some more summary statistics about median time delay, for example.} 
%Nevertheless, the three time delay distributions shown in Figure~\ref{fig:CDF_PLtimedelay_varyZ} are all consistent with one another within the current observational uncertainties, implying that the assumed metallicity threshold does not strongly impact the conclusions about the time delay distribution. 

In addition to varying $Z_\mathrm{thresh}$, we also explore how different assumptions about the metallicity scatter $\sigma_{\log Z}$ affect our inference about the time delay distribution. For a $p(\tau) \propto \tau^{-1}$ time delay distribution, we show the posterior on the minimum time delay for different values of $Z_\mathrm{thresh}$ and $\sigma_{\log Z}$ in Figure~\ref{fig:ptau-min_metallicity}.
For all $Z_\mathrm{thresh}$ and $\sigma_{\log Z}$, we infer a preference for small time delays relative to our prior, with the posterior on $\tau_\mathrm{min}$ peaking at 10 Myr, the smallest time delay in our prior. 
Unsurprisingly, the preference for small time delays is strongest when we assume that the progenitor rate follows the total SFR without any metallicity dependence. Under this assumption, we find $\tau_\mathrm{min} < 2.2$ Gyr (90\% upper limit).
For $\sigma_{\log Z} = 0.4$ dex, we find $\tau_\mathrm{min} < 3.0$ Gyr for $Z_\mathrm{thresh} = 0.3\,Z_\odot$ and $\tau_\mathrm{min} < 3.8$ Gyr for $Z_\mathrm{thresh} = 0.1\,Z_\odot$, following the trends seen in the bottom left panel of Figure~\ref{fig:kappa_timedelay}.
When we assume that the metallicity distribution at each redshift is relatively broad ($\sigma_{\log Z} = 0.6$ dex), shown by the dashed lines of Figure~\ref{fig:ptau-min_metallicity}, the inferred time delay distribution is less sensitive to $Z_\mathrm{thresh}$, with $\tau_\mathrm{min} < 2.9$ Gyr even for the strictest metallicity threshold that we consider, $Z_\mathrm{thresh} = 0.1\,Z_\odot$.

Although we use the mean metallicity-redshift relation of Eq.~\ref{eq:Z-z} from~\citet{2017ApJ...840...39M} throughout this subsection, adopting a different mean metallicity-redshift relation does not significantly affect the conclusions compared to current GW uncertainties on the inferred merger rate. For example, if we instead adopt the mean metallicity-redshift relation from~\citet{2006ApJ...638L..63L}, the values for the merger rate evolution slope $\kappa$ in the bottom left panel of Figure~\ref{fig:kappa_timedelay} increase by a nearly uniform amount of $\sim 0.3$ across the plotted values of $\tau_\mathrm{min}$ and $Z_\mathrm{thresh}$. 

\subsection{Effect of black hole mass}
\label{sec:mass}
\begin{figure}
    \centering
    \includegraphics[width = 0.5\textwidth]{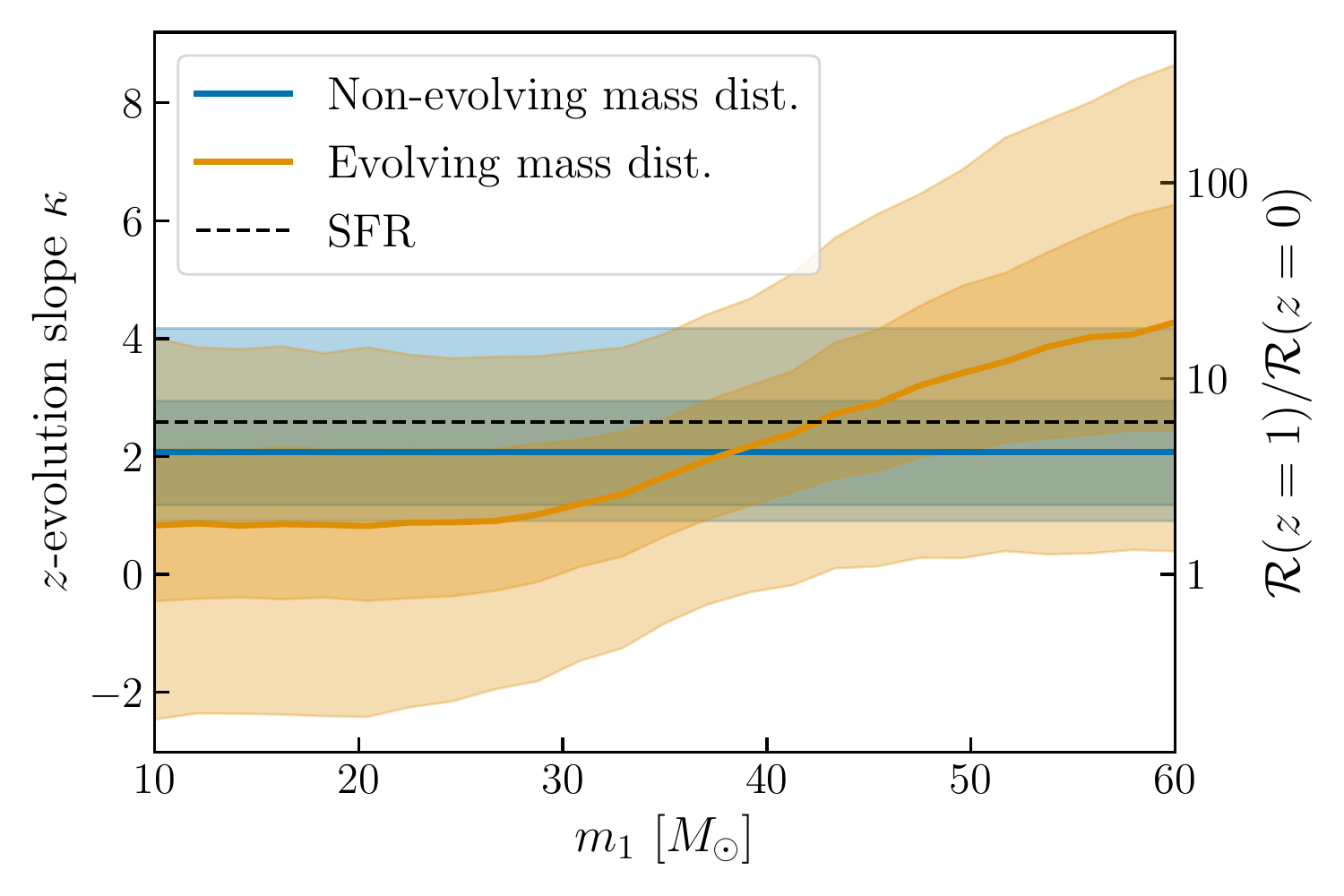}
    \caption{Inferred ratio of the merger rate at $z =1$ to $z = 0$, or approximate power-law slope $\kappa = \frac{\ln (\mathcal{R}_1/ \mathcal{R}_0)} {\ln 2}$, for BBH systems of a given primary mass. The blue band corresponds to the model of Section~\ref{sec:z-evol} in which the mass distribution is independent of redshift. The orange band corresponds to a model in which the mass distribution can evolve with redshift, from~\citet{2021arXiv210107699F}. Solid lines denote medians and shaded bands denote central 50\% and 90\% credible intervals. The dashed black line shows the rate evolution of the SFR.}
    \label{fig:R1R0_mass}
\end{figure}

In the previous subsections, we have assumed that the BBH merger rate evolves with redshift independently of mass, so that if we consider the merger rate within different BBH mass bins, the ratio of the merger rate at $z = 1$ to $z = 0$, $\mathcal{R}_1/ \mathcal{R}_0$, is the same at all masses (shown by the blue band of Figure~\ref{fig:R1R0_mass}). This assumption is supported by population analyses of GWTC-2, which do not find strong evidence that the BBH mass distribution evolves with redshift~\citep{2021arXiv210107699F}. Nevertheless, BBH systems may experience different evolutionary processes, including different metallicity dependences and time delays, based on their masses. 
In fact, BBH systems in different mass ranges might be produced by different formation channels entirely~\citep{2021ApJ...910..152Z}. 

Fitting a population model that allows the mass distribution to evolve with redshift, \citet{2021arXiv210107699F} found a mild preference that the rate increases more steeply from $z = 0$ to $z = 1$ for heavier BBH systems compared to lighter systems. The orange band of Figure~\ref{fig:R1R0_mass} shows $\kappa = \frac{\ln (\mathcal{R}_1/ \mathcal{R}_0)} {\ln 2}$ as a function of primary mass, inferred using the evolving broken power-law model of~\citet{2021arXiv210107699F}. We can use these results to infer different time delay distributions and/or progenitor formation rates as a function of BBH mass. 

If we assume that the progenitor formation rate follows the same SFR across all masses but different time delay distributions, the high mass ($m_1 \sim 50\,M_\odot$) BBH systems exhibit a marginally stronger preference for small time delays than the low mass ($m_1 \sim 15\,M_\odot$) BBH systems. Fitting for the minimum time delay in a $\tau^{-1}$ distribution, we find $\tau_\mathrm{min} < 3.2$ Gyr for BBH systems with $m_1 = 15\,M_\odot$ and  $\tau_\mathrm{min} < 2.4$ Gyr for BBH systems with $m_1 = 50\,M_\odot$ (90\% upper limits); in other words, the two posteriors are completely consistent with one another.

Alternatively, it is possible, although not required by the data, that the progenitor formation rate of the high-mass BBH systems peaks at earlier redshifts, perhaps because of a stricter requirement for low metallicities. However, at all threshold metallicities $Z_\mathrm{thresh} \geq 0.1\,Z_\odot$, in order for the merger rate to evolve faster than $\kappa \sim 2$, we require time delay distributions that are steeper than $\alpha_\tau = -1$ according to the bottom left panel of Figure~\ref{fig:kappa_timedelay}. We also note that high mass BBH mergers may be hierarchical merger products of lower mass BBHs, which would create a complicated dependence of the merger rate evolution between different mass bins (see \citealt{2021arXiv210503439G} for a review).

\section{Progenitor formation rate}
\label{sec:formation}

\begin{figure}
    \centering
    \includegraphics[width = 0.5\textwidth]{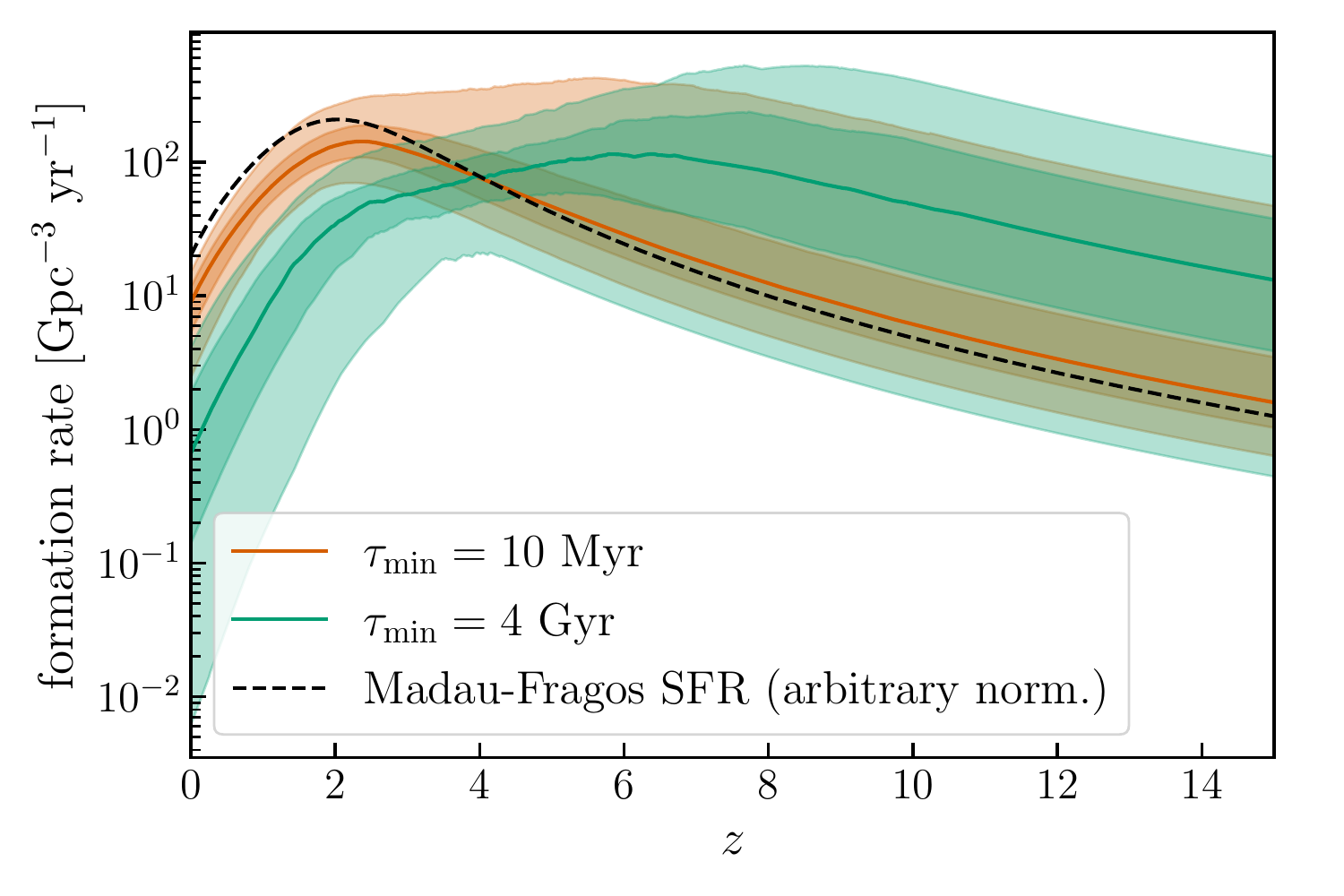}
    \caption{Inferred progenitor formation rate for a fixed power-law time delay model with $\alpha_\tau = -1$. We consider two different minimum delay times: $\tau_\mathrm{min} = 10$ Myr (orange) and $\tau_\mathrm{min} = 4$ Gyr (green). The progenitor formation rate is assumed to follow the low-metallicity SFR where we fit for the threshold metallicity, the scatter around the mean metallicity relation, and the formation efficiency. Solid lines show medians and shaded bands show central 50\% and 90\% credible regions. For reference, the dashed black line shows the Madau-Fragos SFR (Eq.~\ref{eq:sfr}) with an arbitrary normalization.}
    \label{fig:formationrate_metallicity}
\end{figure}

While in the previous section we fixed the progenitor formation rate and extracted the time delay distribution, in this section we consider the reverse problem and extract the progenitor formation rate for a fixed time delay distribution. We use the SFR of Eq.~\ref{eq:sfr} and the mean metallicity-redshift relation of Eq.~\ref{eq:Z-z}~\citep{2017ApJ...840...39M}, and fit for the threshold metallicity $Z_\mathrm{thresh}$, the scatter around the mean metallicity-redshift relation $\sigma_{\log Z}$, and the normalization $R_{f, 0}$. The inferred progenitor formation rates for two different $\tau^{-1}$ time delay models are shown in Figure~\ref{fig:formationrate_metallicity}. We find that a time delay distribution that favors long time delays ($\tau_\mathrm{min} = 4$ Gyr) is possible only if the progenitor rate peaks at \replaced{$z = 5.4^{+3.0}_{-3.2}$}{$z > 3.9$, or $t_L > 12$ Gyr (90\% credibility, reweighting to a flat prior on the peak redshift in the range 2.1--9.)}. \added{This result is driven by the absence of a peak in the BBH merger rate density at $z < 1$ ($t_L \lesssim 8$ Gyr), so that for any minimum delay time $\tau_\mathrm{min}$, the progenitor formation rate probably peaked at a lookback time $t_L > \tau_\mathrm{min} + 8$ Gyr.} A time delay distribution with $\tau_\mathrm{min} = 4$ Gyr requires a strict metallicity threshold, favoring the lowest metallicity threshold in our prior ($Z_\mathrm{thresh} = 0.1\,Z_\odot$) by a Bayes factor of 40 compared to no metallicity weighting; see also the bottom left panel of Figure~\ref{fig:kappa_timedelay}. 

Meanwhile, for a time delay distribution that favors short time delays ($\tau_\mathrm{min} = 10$ Myr), as predicted for the isolated binary evolution channel~\citep{2010ApJ...716..615O,2012ApJ...759...52D,2017MNRAS.472.2422M,2019MNRAS.490.3740N}, we infer a progenitor formation rate that matches the SFR well, but also permits the peak formation rate to occur at higher redshifts. Assuming the short-time delay model, the data remain uninformative about the metallicity dependence, with our posterior over $Z_\mathrm{thresh}$ recovering the prior. Also interesting to note is that the inferred amplitude of the progenitor formation rate at $z = 0$, $R_{f, 0}$, related to the BBH formation efficiency, differs for the two time delay models, with $R_{f,0} = 0.74^{+3.65}_{-0.73}\,\mathrm{Gpc}^{-3}\,\mathrm{yr}^{-1}$ for the $\tau_\mathrm{min} = 4$ Gyr model and $R_{f,0} = 7.8^{+8.3}_{-5.0}\,\mathrm{Gpc}^{-3}\,\mathrm{yr}^{-1}$ for the $\tau_\mathrm{min} = 10$ Myr model.

\section{Conclusion}
\label{sec:conclusion}
We have derived the first observational constraints on the BBH time delay distribution and the formation rate of their progenitors from the latest LIGO-Virgo catalog, GWTC-2. We found that with only 44 BBH events out to $z \lesssim 1$, we can already rule out models in which time delays are longer than $\sim 3$ Gyr if the progenitor formation rate is close to the SFR. Our main results are as follows:   
\begin{enumerate}
    \item \textbf{Short time delays are favored for all progenitor formation rates we consider.} For a progenitor formation rate that follows the SFR, we find that 43\%-100\% of mergers have time delays $\tau < 4.5$ Gyr. 
    For a $p(\tau) \propto \tau^{-1}$ time delay distribution, we find that the time delay distribution peaks below $\tau_\mathrm{min} < 2.2$ Gyr if the progenitor formation rate follows the SFR. This corresponds to median time delays $\tau_{50\%} = 2.8^{+3.3}_{-2.6}$ Gyr. If the progenitor formation rate follows the low-metallicity ($Z < 0.3\,Z_\odot$) SFR, we find $\tau_\mathrm{min} < 3.0$ Gyr, with $\tau_\mathrm{50\%} = 3.9^{+3.2}_{-3.5}$ Gyr.
    \item \textbf{If the time delay distribution favors longer delay times ($\tau \gtrsim 4$ Gyr), the progenitor formation rate must peak earlier than the SFR.} For example, for a $\tau^{-1}$ time delay distribution with $\tau_\mathrm{min} = 4$ Gyr, the progenitor formation rate peaks at \replaced{$z = 5.4^{+3.0}_{-3.2}$}{$z > 3.9$}, and is fairly low at $z = 0$, with $R_{f, 0} = 0.74^{+3.65}_{-0.73}$ Gpc$^{-3}$ yr$^{-1}$. If the progenitor formation rate is related to the SFR, this requires that the progenitors only form at low metallicities $Z < 0.1\,Z_\odot$. On the other hand, if we assume a $\tau^{-1}$ time delay distribution with $\tau_\mathrm{min} = 10$ Myr, motivated by predictions from isolated binary evolution, we constrain the progenitor formation rate at $z = 0$ to be $R_{f, 0} = 7.8^{+8.3}_{-5.0}$ Gpc$^{-3}$ yr$^{-1}$. The shape of the progenitor formation rate matches the SFR well. 
    \item \textbf{There is no strong evidence that the time delay distribution or the progenitor formation rate varies with mass.} However, it is possible that more massive systems experience shorter delay times and/or a stricter low-metallicity threshold. 
\end{enumerate}

\added{We can use the constraints on the time delay distribution to probe the evolutionary pathways that give rise to BBH systems, with the caveat that our analysis only applies to formation channels in which BBH progenitors follow the (low-metallicty) star-formation rate. Thus, our measurement of the time delay distribution does not directly apply to primordial BH channels (see \citealt{2018CQGra..35f3001S} for a review), BBH mergers in AGN disks, and dynamical assembly in globular clusters if they were predominantly formed during the reionization epoch, rather than concurrently with star formation~\citep{2018RSPSA.47470616F,2020arXiv201114541R}.}

With the current GW catalog, the inferred time delay distribution remain consistent with most of the formation channels discussed in Section~\ref{sec:intro}. There may be hints of tension with formation scenarios that favor longer time delays, like stellar triples (where most time delays are greater than $\sim 1$ Gyr, and BBH formation is expected to be relatively efficient at high metallicities), and chemically homogeneous binaries (where most time delays are greater than a few Gyr, but the low-metallicity requirement for BBH formation may be much stricter). 
%\added{It is also important to note that our constraints on the time delay distribution only apply to formation channels in which the progenitors follow the (low-metallicity) star-formation rate.}

Nevertheless, we can see from Figure~\ref{fig:kappa_timedelay} how our constraints on the time delay distribution and the progenitor metallicity dependence will improve as the measurement of the redshift evolution slope $\kappa$ tightens. At O3 sensitivity, the width of the 90\% credible interval on $\kappa$ converges with the number of events $N$ as $\Delta_{90\%}(\kappa) \sim 31/\sqrt{N}$~\citep{2018ApJ...863L..41F}; this is well matched by the current measurement $\Delta_{90\%}(\kappa) = 4.3$ with $\sim 50$ events. With more sensitive detectors, the measurement of $\kappa$ is expected to converge faster; $\Delta_{90\%}(\kappa)$ scales inversely with the average $\ln(1+z)$ among detected events. For detectors that are 50\% more sensitive, as expected for Advanced LIGO at design sensitivity, it is likely that 500 events (around a year of observation) will constrain $\kappa$ to $\Delta_{90\%}(\kappa) = 1$. If the inferred merger rate prefers relatively steep evolution ($\kappa \gtrsim 2$), this will put pressure on many time delay models, requiring time delay distributions that are steeper than flat-in-log and/or a strict low-metallicity requirement for progenitor formation (see Figure~\ref{fig:kappa_timedelay}). In addition to inferring the time delay distribution and the BBH formation efficiency as a function of metallicity, the BBH merger rate can yield insight into the metallicity evolution of the universe. Another exciting application of this calculation would be to determine which time delay model (for example, power law versus log-normal) best fits the data, which would reveal details of the BBH formation model. The binned time delay model of Section~\ref{sec:binned} has the advantage of naturally incorporating all possible models via its flexibility, although meaningful model selection will likely only be possible once the peak of the merger rate is resolved, either with continued observation of the stochastic background~\citep{2020ApJ...896L..32C} or with the next generation of ground-based GW detectors~\citep{2019ApJ...886L...1V}.

\acknowledgments
MF is supported by NASA through NASA Hubble Fellowship grant HST-HF2-51455.001-A awarded by the Space Telescope Science Institute. VK is supported by a CIFAR G+EU Senior Fellowship and Northwestern University.
This research has made use of GWTC-2 parameter estimation samples obtained from the Gravitational Wave Open Science Center (https://www.gw-openscience.org/ ), a service of LIGO Laboratory, the LIGO Scientific Collaboration and the Virgo Collaboration. LIGO Laboratory and Advanced LIGO are funded by the United States National Science Foundation (NSF) as well as the Science and Technology Facilities Council (STFC) of the United Kingdom, the Max-Planck-Society (MPS), and the State of Niedersachsen/Germany for support of the construction of Advanced LIGO and construction and operation of the GEO600 detector. Additional support for Advanced LIGO was provided by the Australian Research Council. Virgo is funded, through the European Gravitational Observatory (EGO), by the French Centre National de Recherche Scientifique (CNRS), the Italian Istituto Nazionale di Fisica Nucleare (INFN) and the Dutch Nikhef, with contributions by institutions from Belgium, Germany, Greece, Hungary, Ireland, Japan, Monaco, Poland, Portugal, Spain. 
This document has been assigned LIGO document number P2100164.
\software{
\textsc{astropy}~\citep{2018AJ....156..123A},
\textsc{emcee}~\citep{2013ascl.soft03002F},
\textsc{NumPy}~\citep{2020Natur.585..357H},
\textsc{Matplotlib}~\citep{2007CSE.....9...90H},
\textsc{PESummary}~\citep{2020arXiv200606639H},
\textsc{PyMC3}~\citep{2016ascl.soft10016S},
\textsc{SciPy}~\citep{2020zndo...4100507V},
\textsc{seaborn}~\citep{2020ascl.soft12015W},
\textsc{Theano}~\citep{2016arXiv160502688T}.
}

\bibliography{references}{}
\bibliographystyle{aasjournal}

\appendix
\section{Statistical framework}
\label{sec:methods}

We write the differential number density of BBH systems as:
\begin{equation}
    \frac{d\mathcal{N}}{dm_1dm_2d\chi_\mathrm{eff}dz dt_d} \equiv N p(m_1, m_2, \chi_\mathrm{eff}, z),
\end{equation}
where $m_1$ and $m_2$ are the primary and secondary component masses, $\chi_\mathrm{eff}$ is the effective inspiral spin, $z$ is the source redshift, $t_d$ is time measured in the detector-frame, $p$ is a normalized probability density that integrates to unity over the considered range of masses, spins, and redshifts, and $N$ is the number of BBHs within the mass, spin, and redshift range that merge during the observing time $T_\mathrm{obs} = \int dt_d$. We are primarily concerned with the redshift distribution in this work, but the inferred redshift distribution correlates with the inferred mass and, to a lesser extent, $\chi_\mathrm{eff}$ distribution, and so we must consider these properties jointly. 
The merger rate density is:
\begin{align}
\label{eq:rate-z}
\mathcal{R}(z) &\equiv  \frac{d\mathcal{N}}{dV_c dt_s} = \frac{d\mathcal{N}}{dz dt_d}\left(\frac{dV_c}{dz}\right)^{-1}\frac{dt_d}{dt_s} \nonumber \\
&= \frac{d\mathcal{N}}{dz} \left(\frac{dV_c}{dz}\right)^{-1} \frac{T_\mathrm{obs}}{1 + z},
\end{align}
where:
\begin{equation}
    \frac{d\mathcal{N}}{dz} \equiv N \int p(m_1, m_2, \chi_\mathrm{eff}, z) dm_1 dm_2 d\chi_\mathrm{eff}.
\end{equation}
We adopt a parameterized model to describe the population distribution $p(m_1, m_2, \chi_\mathrm{eff}, z \mid \lambda)$, where $\lambda$ are the parameters of the model. For now, we assume that the mass and spin distributions are independent of redshift, so that:
\begin{equation}
\label{eq:lambda}
    p(m_1, m_2, \chi_\mathrm{eff}, z \mid \lambda) = p(m_1, m_2 \mid \lambda_m)p(\chi_\mathrm{eff} \mid \lambda_\chi)p(z \mid \lambda_z),
\end{equation}
and:
\begin{equation}
    \mathcal{R}(z) = Np(z)\left(\frac{dV_c}{dz}\right)^{-1} \frac{T_\mathrm{obs}}{1 + z}.
\end{equation}
We use the \textsc{Broken power law} mass model from~\citet{2020arXiv201014533T} to describe $p(m_1, m_2 \mid \lambda_m)$ and the \textsc{Gaussian} spin model from~\citet{2020ApJ...895..128M,2020arXiv201014533T} to describe $p(\chi_\mathrm{eff} \mid \lambda_\chi)$. The primary mass distribution follows a power law with slope $\alpha_1$ between $m_\mathrm{min}$ and $m_\mathrm{break}$ and slope $\alpha_2$ between $m_\mathrm{break}$ and $m_\mathrm{max}$. The mass ratio distribution follows a power law with slope $\beta_q$. The $\chi_\mathrm{eff}$ distribution is described by a Gaussian with mean $\mu_\mathrm{\chi_\mathrm{eff}}$ and standard deviation $\sigma_\mathrm{\chi_\mathrm{eff}}$, truncated to the physical range $-1 < \chi_\mathrm{eff} < 1$~\citep{2019MNRAS.484.4216R,2020ApJ...895..128M}. For the redshift evolution model $p(z \mid \lambda_z)$, we write:
\begin{equation}
\label{eq:pz}
    p(z \mid \lambda_z) \propto \frac{dV_c}{dz}\frac{1}{1+z}f(z \mid \lambda_z).
\end{equation}
For $f(z)$, we include the possibility that the merger rate peaks at a redshift $z < 1$ by using a smoothly broken power law in $(1 + z)$:
where~\citep{2018AJ....156..123A}:
\begin{equation}
\label{eq:pz-sbpl}
    f(z \mid \kappa, \gamma, z_p, \Delta) = \left(\frac{1 + z}{1 + z_p}\right)^\kappa \left[\frac{1+ \left(\frac{1 + z}{1 + z_p} \right)^{1/\Delta}}{2} \right]^{(\gamma - \kappa) \Delta},
\end{equation}
and we fix the smoothing parameter $\Delta = 0.2$. This model with $\kappa = 2.6$ $\gamma = -3.6$, $z_p = 2.2$ and $\Delta = 0.16$ can reproduce the shape of the~\citet{2017ApJ...840...39M} SFR. We note that with the current data, we cannot observe the peak redshift $z_p$, but can rule out that the merger rate peaks at $z \lesssim 1$. 

\begin{table}[t]
    \centering
    \begin{tabular}{ c  p{11cm} p{2mm} p{3cm} }
        \hline
        {\bf Parameter} & \textbf{Description} &  & \textbf{Prior} \\\hline\hline
        $m_\mathrm{min} / M_\odot$ & Minimum BH mass &  & U(2, 10)\\
        $m_\mathrm{break} / M_\odot$ & Mass at which the power law describing the primary mass distribution breaks &  & U(20, 65)\\
        $m_\mathrm{max} / M_\odot$ &  Maximum BH mass &  & U(65, 100)\\
        $\alpha_1$ & Power-law slope of the primary mass distribution for masses below $m_\mathrm{break}$ &  & U(-5, 2) \\
        $\alpha_2$ & Power-law slope of the primary mass distribution for masses above $m_\mathrm{break}$ &  & U(-12, 2) \\
        $\beta_q$ & Power-law slope of the mass ratio distribution &  & U($-4$, $12$) \\
        \hline
        $\mu_{\chi_\mathrm{eff}}$ & Mean of the Gaussian describing the $\chi_\mathrm{eff}$ distribution & & U(-0.5, 0.5) \\
        $\sigma{\chi_\mathrm{eff}}$ & Standard deviation of the Gaussian describing the $\chi_\mathrm{eff}$ distribution & & U(0.02, 1) \\ 
        \hline
        $z_p$ & Redshift at which the merger rate peaks & & U(0, 3) \\
        $\kappa$ & Power-law slope in $(1 + z)$ of the merger rate evolution for $z < z_p$ & & U(0, 6) \\
        $\gamma$ &  Power-law slope in $(1 + z)$ of the merger rate evolution for $z > z_p$ & & U(-6, 0) \\
        \hline
    \end{tabular}
    \caption{
    \added{Summary of population hyperparameters $\lambda$ in Eq.~\ref{eq:lambda}. We group the hyperparameters into three groups: the mass distribution parameters $\lambda_m$, the spin distribution parameters $\lambda_\chi$, and the redshift distribution parameters $\lambda_z$. The notation U$(a, b)$ denotes a uniform distribution between $a$ and $b$.}
    }
  \label{tab:priors}
\end{table}

\begin{table}[t]
    \centering
    \begin{tabular}{ c  p{11cm} p{2mm} p{3cm} }
        \hline
        {\bf Parameter} & \textbf{Description} &  & \textbf{Prior} \\\hline\hline
        $\lbrace p_i(b_{i+1} - b_i) \rbrace_{i = 1, 2, 3}$ & Fraction of systems with delay times between $b_i$ and $b_{i + 1}$ in the binned time delay model  & & Dir(0.5, 0.5, 0.5)\\
        $\tau_\mathrm{min} / \mathrm{Gyr}$ & Minimum time delay in the power-law time delay model & & U(0.01, 5) \\
        $\alpha_\tau$ & Power-law slope of the time delay distribution in the power-law model & & U(-3, 1)\\
        $R_{f,0} /\, \mathrm{Gpc}^{-3}\,\mathrm{yr}^{-1}$ & Progenitor formation rate at $z = 0$ &  & LU($10^{-3}$, 100)\\
        $Z_\mathrm{thresh}/ Z_\odot$ & Threshold metallicity for progenitor formation &  & U(0.1, 2)\\
        $\sigma_{\log Z}$ & Scatter about the mean metallicity-redshift relation &  & U(0.2, 0.6)\\
        \hline
    \end{tabular}
    \caption{
    \added{Summary of hyperparameters describing the time delay distribution and the progenitor formation rate. The notation Dir$(\alpha_1, \ldots, \alpha_k)$ denotes a Dirichlet distribution with concentration parameters $\alpha_i$,  U$(a, b)$ a uniform distribution between $a$ and $b$, and LU$(a, b)$ a log-uniform distribution, so that $X \sim \mathrm{LU}(a, b)$ implies $\log X \sim \mathrm{U} (\log a, \log b)$.}
    }
  \label{tab:prior-theta}
\end{table}

We fit for the population parameters $\lambda$ with the usual hierarchical Bayesian framework~\citep{2010PhRvD..81h4029M,2019MNRAS.486.1086M,2019RNAAS...3...66F}, using the same parameter estimation samples and detector sensitivity estimate as~\citet{2020arXiv201014533T}. \added{We take broad, flat priors on all parameters $\lambda$ as detailed in Table~\ref{tab:priors}}, and a flat-in-log prior on the normalization parameter $N$. \deleted{For the mass distribution parameters that make up $\lambda_m$, we take: $m_\mathrm{min} /M_\odot \in \left[2, 10 \right]$, $m_\mathrm{break} / M_\odot \in \left[20, 65 \right]$, $m_\mathrm{max} / M_\odot \in \left[65, 100 \right]$, $\alpha_1 \in \left[-5, 2\right]$, $\alpha_2  \in \left[-12, 2 \right]$ and $\beta_q \in \left[ -4, 12\right]$. For the spin distribution parameters in $\lambda_\chi$, we take $\mu_{\chi_\mathrm{eff}} \in \left[-0.5, 0.5\right]$, $\sigma_{\chi_\mathrm{eff}} \in \left[0.02, 1 \right]$. For the redshift evolution parameters in $\lambda_z$, we take $\kappa \in \left[-6, 0\right]$, $z_p \in \left[0 , 3\right]$, and $\gamma \in \left[-6, 0\right]$, noting that with the current data, we cannot observe the peak redshift $z_p$, but can rule out that the merger rate peaks at $z \lesssim 1$.} We sample from the posterior over the model parameters $\lambda$ with PyMC3~\citep{2016ascl.soft10016S}.

Our goal is to extract information about the delay time distribution and the progenitor formation rate from this inferred merger rate evolution.
Given the progenitor formation rate $R_f$ and a time delay distribution $p(\tau)$, we can calculate the resulting merger rate as a function of lookback time: 
\begin{equation}
\label{eq:rate-integral}
    \mathcal{R}(t_L) = \int_{\tau_\mathrm{min}}^{\tau_\mathrm{max}} R_f(t_L + \tau)p(\tau)d\tau,
\end{equation}
where $t_L$ is the lookback time corresponding to merger and $t_L + \tau$ is the lookback time corresponding to formation for a given time delay $\tau$.
Under an assumed cosmological model, the lookback time can be calculated from the redshift $z$. We use the median \emph{Planck 2015} cosmological parameters~\citep{2016A&A...594A..13P}.
Throughout this work, we assume the earliest progenitor formation time was 13.5 Gyr ago, corresponding to a maximum formation redshift $z = 14$, so we fix $\tau_\mathrm{max} = 13.5$ Gyr to restrict to systems that have already merged. 

If we have a binned time delay distribution, as in Eq.~\ref{eq:binned},
\begin{equation}
    p(\tau) = \sum_{i = 1}^{n} p_i \Theta(b_i \leq \tau < b_{i+1}),
\end{equation}
it is straightforward to compute the merger rate of Eq.~\ref{eq:rate-integral}:
\begin{equation}
\label{eq:merger-rate-binned}
    \mathcal{R}(t_L) = \sum_{i = 1}^n p_i \left(F(t_L + b_{i+1}) - F(t_L + b_i)\right),
\end{equation}
where $F$ is the integral of the formation rate as a function of lookback time, $F(x) = \int_0^x R_f(t_L) dt_L$. We convert lookback time to redshift, $z(t_L)$, to write Eq.~\ref{eq:merger-rate-binned} in terms of $\mathcal{R}(z)$.

We assume that the progenitor formation rate $R_f$ depends on the SFR, as reported in~\citet{2017ApJ...840...39M}:
\begin{equation}
\label{eq:sfr}
    R_\mathrm{SFR}(z) \propto \frac{\left(1+z\right)^{2.6}}{1+ \left(\frac{1 + z}{3.2}\right)^{6.2}}.
\end{equation}
 We also consider models in which the formation rate of BBH progenitors does not follow the \emph{total} SFR, but rather follows only the \emph{low-metallicity} SFR below some threshold metallicity $Z_\mathrm{thresh}$. This assumption is equivalent to a model in which the BBH formation efficiency depends on metallicity, and this metallicity dependence can be approximated by a step function: the efficiency is constant for $Z \leq Z_\mathrm{thresh}$ and sharply turns off at $Z > Z_\mathrm{thresh}$. When considering the low-metallicity SFR, we adopt the mean metallicity-redshift relation from~\citet{2017ApJ...840...39M}:
\begin{equation}
\label{eq:Z-z}
    \langle \log_{10} Z(z) / Z_\odot \rangle = 0.153 - 0.074z^{1.34}
\end{equation}
The distribution of metallicities at each redshift, particularly the scatter, is uncertain~\citep{2019MNRAS.482.5012C}. We assume that $\log_{10}Z(z)$ follows a normal distribution at each $z$ with some standard deviation $\sigma_{\log Z}$. We adopt the default value of $\sigma_{\log Z} = 0.4$ dex, as in~\citet{2019MNRAS.490.3740N} and \citet{2021arXiv210212495B}, although we sometimes treat it as a free parameter in the model.

For the binned time delay model, we substitute $\mathcal{R}(z)$ as calculated in Eq.~\ref{eq:merger-rate-binned} for $f(z \mid \lambda_z)$ in Eq.~\ref{eq:pz}, and jointly fit $\lambda_m$, $\lambda_\chi$, and the time delay bin heights $p_i$, as described in Section~\ref{sec:binned} of the main text. \added{See Tables~\ref{tab:priors} and \ref{tab:prior-theta} for the sampling priors on the hyperparameters.}
When fitting the other time delay and progenitor formation rate models, we approximate the full hierarchical Bayesian likelihood as follows.

We observe that the joint posterior for $\mathcal{R}_1$ and $\mathcal{R}_0$ is insensitive to the assumed parameterization of the redshift distribution, whether we fit a one-parameter power law in $(1 + z)$, the smoothly broken power-law model of Eq.~\ref{eq:pz-sbpl}, or the physical, binned time delay model of the previous subsection~\ref{sec:binned}. 
The marginal 1-dimensional posteriors are $\mathcal{R}_0 = 18^{+17}_{-9}$ Gpc$^{-3}$ yr$^{-1}$ and $\mathcal{R}_1 = 66^{+190}_{-46}$ Gpc$^{-3}$ yr$^{-1}$.
This observation allows us to speed up the inference by approximating the likelihood of the GW catalog given the time delay distribution as follows. We are interested in fitting for the parameters specifying the time delay distribution --- for example, a power law with slope $\alpha_\tau$ and minimum delay $\tau_\mathrm{min}$ --- together with the parameters specifying the formation rate: the metallicity threshold $Z_\mathrm{thresh}$, the scatter about the mean-metallicity relation $\sigma_{\log Z}$, and the amplitude of the formation rate $R_f(0) = R_{f,0}$ (equivalently, the BBH formation efficiency). Denoting these parameters by $\theta \equiv \lbrace  \alpha_\tau, \tau_\mathrm{min}, Z_\mathrm{thresh}, \sigma_{\log Z}, R_{f,0} \rbrace$, we calculate $\mathcal{R}_1(\theta)$ and $\mathcal{R}_0(\theta)$ by Eq.~\ref{eq:rate-integral} as in Figure~\ref{fig:kappa_timedelay}. We are interested in the posterior probability distribution of $\theta$ given the GW data, $d$:
\begin{equation}
\label{eq:theta-post}
    p(\theta \mid d) \propto p(d \mid \theta)p_0(\theta),
\end{equation}
where $p_0$ represents the prior probability and we approximate the likelihood:
\begin{equation}
\label{eq:theta-approx-l}
p(d \mid \theta) \approx p(d \mid \mathcal{R}_1(\theta), \mathcal{R}_0(\theta)). 
\end{equation}
We calculate $p(d \mid \mathcal{R}_1, \mathcal{R}_0)$ from the phenomenological population fit of Section~\ref{sec:z-evol}, where we computed the likelihood given the population parameters $p(d \mid N, \lambda_m, \lambda_\chi, \lambda_z)$. From the population parameters $N$ (1-dimensional) and $\lambda_z$ (3-dimensional), we calculate $\mathcal{R}_0(N, \lambda_z)$ and $\mathcal{R}_1(N, \lambda_z)$ according to Eq.~\ref{eq:rate-z}. Given $(N, \lambda)$ posterior samples drawn from:
\begin{equation}
    p(N, \lambda \mid d) \propto p(d \mid N, \lambda)p_0(N, \lambda),
\end{equation}
where $\lambda \equiv \lbrace \lambda_m, \lambda_\chi, \lambda_z \rbrace$,
we apply the function $\mathcal{R}(z, \lambda_z, N)$ at $z = 0$ and $z = 1$ to get samples from the probability density $p(\mathcal{R}_0, \mathcal{R}_1 \mid d)$. We then also draw prior samples $(N, \lambda)$ drawn from $p_0(N, \lambda)$ to calculate $p_0(\mathcal{R}_0, \mathcal{R}_1)$. Given these two sets of posterior and prior samples, we approximate both the posterior density and the prior density with a Gaussian kernel density estimate to evaluate the approximate likelihood:
\begin{equation}
p(d \mid \mathcal{R}_0, \mathcal{R}_1) \propto \frac{ p(\mathcal{R}_0, \mathcal{R}_1 \mid d)}{p_0(\mathcal{R}_0, \mathcal{R}_1)},
\end{equation}
which we can substitute into the desired posterior probability distribution for $\theta$ through Eqs.~\ref{eq:theta-post} and~\ref{eq:theta-approx-l}. \replaced{For the prior $p_0(\theta)$, our default prior is flat on $\tau_\mathrm{min} / \mathrm{Gyr} \in \left[0.01, 5 \right]$, $\alpha_\tau \in \left[-3, 1\right]$, $\log_{10}(R_{f,0} \times \mathrm{Gpc}^3\, \mathrm{yr}) \in \left[-3, 2\right]$, $Z_\mathrm{thresh}/Z_\odot \in \left[0.1, 2\right]$, $\sigma_{\log Z} \in \left[0.2, 0.6\right]$}{For $p_0(\theta)$, our default priors are shown in Table~\ref{tab:prior-theta}}.

\end{document}